\documentclass[preprint]{elsarticle}
\usepackage{lineno,hyperref}
\modulolinenumbers[5]
\usepackage{amsmath,amssymb,amsfonts}
\usepackage{graphicx}
\usepackage{textcomp}
\usepackage{xcolor}
\usepackage{adjustbox}

\newif\ifcolorred
\colorredfalse
\ifcolorred
\newcommand{\red}[1]{\textcolor{red}{#1}}
\else
\newcommand{\red}[1]{\textcolor{black}{#1}}
\fi

\usepackage{pgfplots}
\pgfplotsset{compat=1.8}
\usepgfplotslibrary{statistics}

\usepackage{multirow}

\usepackage{algpseudocode}  

\usepackage{caption}
\usepackage{subcaption}

\journal{Journal of Systems and Software}

\begin{document}

\begin{frontmatter}
\title{Improved Retrieval of Programming Solutions With Code Examples Using a Multi-featured Score}

\author[ufu]{Rodrigo F. Silva}
\ead{rodrigofernandes@ufu.br}

\author[dal]{Mohammad Masudur Rahman} 
\ead{masud.rahman@dal.ca}

\author[ufu,iftm]{Carlos Eduardo Dantas} 
\ead{carlos.dantas@ufu.br}

\author[usask]{Chanchal Roy} 
\ead{chanchal.roy@usask.ca}

\author[epl]{Foutse Khomh}
\ead{foutse.khomh@polymtl.ca}

\author[ufu]{Marcelo A. Maia}
\ead{marcelo.maia@ufu.br}

\address[ufu]{Faculty of Computing, Federal University of Uberlândia, Brazil} 
\address[dal]{Faculty of Computer Science, Dalhousie University, Canada}
\address[iftm]{Instituto Federal do Triângulo Mineiro, Brazil}
\address[usask]{Department of Computer Science, University of Saskatchewan, Canada}
\address[epl]{École Polytechnique de Montréal, Canada}

\begin{abstract}

Developers often depend on code search engines to obtain solutions for their programming tasks. However, finding an expected solution containing code examples along with their explanations is challenging due to several issues. There is a vocabulary mismatch between the search keywords (the query) and the appropriate solutions. Semantic gap may increase for similar bag of words due to antonyms and negation. Moreover, documents retrieved by search engines might not contain solutions containing both code examples and their explanations. So, we propose CRAR (Crowd Answer Recommender) to circumvent those issues aiming at improving retrieval of relevant answers from Stack Overflow containing not only the expected code examples for the given task but also their explanations. Given a programming task, we investigate the effectiveness of  combining information retrieval techniques along with a set of features to enhance the ranking of important threads (i.e., the units containing questions along with their answers) for the given task and then selects relevant answers contained in those threads, including semantic features, like word embeddings and sentence embeddings, for instance, a Convolutional Neural Network (CNN). CRAR also leverages social aspects of Stack Overflow discussions like popularity to select relevant answers for the tasks. Our experimental evaluation shows that the combination of the different features performs better than each one individually. We also compare the retrieval performance with the state-of-art CROKAGE (Crowd Knowledge Answer Generator), which is also a system aimed at retrieving relevant answers from Stack Overflow. We show that CRAR outperforms CROKAGE  in Mean Reciprocal Rank and Mean Recall with small and medium effect sizes, respectively.

\end{abstract}


\end{frontmatter}

\linenumbers

\section{Introduction} \label{Introduction}

Developers often search for solutions on the web for their programming tasks in the form of code examples and their explanations  \cite{Sadowski2015}, spending up to 20\% of their time in this activity \cite{philip2012, Niu2017}. Although code search engines (e.g., GitHub, Google Code Search) can be used for such a task, finding these solutions for programming tasks is still a challenge because not all code examples returned for a query have the same quality \cite{Ying2014}. Developers need to carefully choose appropriate keywords that represent their programming task (i.e., the query) due to the semantic gap existent between the query and the appropriate code. Notwithstanding, they need to continuously reformulate the queries in order to improve the retrieved solutions~\cite{bajracharya2012analyzing,sadowski2015developers}.  
 Furthermore, other related work that aims at tackling the semantic gap  also presents only  the source code for the task~\cite{6233407,campbell2017nlp2code,rahman2017rack,cambronero2019deep,gu2018deep,sachdev2018retrieval} without their accompanying explanations, which are often very useful for reusing any such  available source code. AnswerBot has the inverse limitation, it produces only textual answers where it could have included  code snippets to provide a more understandable answer for the input query~\cite{xu2017answerbot}.

In order to find solutions that have both code examples and respective contextual explanations, developers usually rely on Q\&A platforms such as Stack Overflow to find solutions for their programming tasks. Stack Overflow is the most popular Q\&A community for developers~\cite{baltes2018sotorrent} where they share their experiences, doubts and solutions about programming topics. The forum contains millions of Q\&A discussions\footnote{18.3M questions and 27.9M answers - October, 2019} comprising the use of a wide range of Application Programming Interfaces (APIs). Thus, it is very likely that the respective task might have already been discussed among the threads (i.e., the units containing questions along with their answers). Moreover, the threads contain a variety of solutions using different APIs in the form of curated code examples followed by their explanations.  

As a result, developers can search for solutions in the Stack Overflow content, \red{either using native Stack Overflow search capability or using a general purpose search service},  to obtain already posted solutions for their tasks with no need to wait for an answer for their questions. 
Unfortunately, similarly to the traditional code search engines, Stack Overflow does not overcome the aforementioned problems \red{(e.g., retrieval of solutions without both code and explanation). The same would happen when developers  search for solutions using general-purpose search services}. That is,  solutions can be incomplete, i.e., without code examples that implement the query intent or without accompanying explanations.

\begin{table}[]
\centering
\caption{CROKAGE results for the query: ``How to resize images in Java?"}
\label{tab:soresultsquery}
\begin{tabular}{cccl}
\hline Rank & Answer ID/Thread ID & Answer Score & Goldset (HIT) \\ 
 1    & 16076530/244164 & 2 & No \\
 2    & 32740879/32739699 & 2 & No \\
 3    & 19898341/5150503 & 2 & No \\
 4    & 9403763/9403717 & 5 & No \\
 5    & 6586119/2305670 & 7 & No \\
 6    & 32758346/32739699 & 2 & No \\
 7    & 6444133/6444042 & 3 & No \\
 8    & 12204680/362360 & 13 & No \\
 9    & 9804943/9804069 & 2 & No \\
 10    & 13892750/13892725 & 8 & No \\
\hline 11    & 6585602/1625137 & 16 & Yes \\ 
 12    & 6585887/991349 & 67 & Yes \\ 
 13    & 244177/244164   & 84 & Yes \\
 20    & 5051429/244164   & 49 & Yes \\ 
 107    & 4528136/244164   & 177 & Yes \\ 

\hline
\end{tabular}
\end{table}

In order to fill this gap between programming tasks and complete programming solutions, i.e., those that contain both the code solution and their respective explanation, two recent works, BIKER~\cite{huang2018api} and CROKAGE~\cite{silva2019recommending, crokage2020}, were proposed. These works tackle the aforementioned problems. They leverage word embeddings~\cite{mikolov2013distributed,le2014distributed} to address the semantic gap between the query terms and the appropriate solutions. Furthermore, their solutions are composed by both code snippets that implement the query intent and their respective explanations. 
\red{On the other hand, compared to general-purpose search services, CROKAGE has been shown to provide more simple and direct answers \cite{crokage2020}. Moreover, search engines deliver a different kind of response to a query, i.e., a list of links,  which  should  be  investigated  by  the  developer  to  find  out  an  appropriate  answer  for  the query.  Thus, general-purpose search engines provide a different type of content, compared to  BIKER,  CROKAGE  and  CRAR,  and  hence,  might  not  be  an  appropriate  baseline  to compare with.}
Besides, that BIKER recommendation did not perform as good as other baselines, including, CROKAGE, as shown by Silva et al \cite{silva2019recommending}. So, we will also not consider BIKER as a baseline to overcome. 
CROKAGE combines word embeddings with Information Retrieval (IR) techniques to overcome BIKER's limitations. It retrieves relevant answers from Stack Overflow whose explanations comprise of multiple APIs. However, there is still room for improvement of CROKAGE effectiveness.  The design space for such kind of retrieval approach is very broad, and deciding how to choose the proper features and how to effectively integrate them is far from being a trivial task. There are some features that CROKAGE disregard that could be investigated to improve the retrieval performance.
A motivating example is the query \textit{``How to resize images in Java"} whose results are shown in Table~\ref{tab:soresultsquery}. There are at least 37 relevant Stack Overflow answers IDs, and CROKAGE did not hit any of them in the top-10 rank. The best ranked CROKAGE answer found has the ID 16076530, and it is in the Thread 244164, but this is not a relevant answer. Nonetheless, that same thread has three (out of 37) relevant answers, with IDs 244177,5951429 and 4528136, where all of them are out of the CROKAGE top-10 answers. Interestingly, those answers have high answer score (upvotes minus downvotes), and CROKAGE's best ranked answer has a low score, and is from that same thread. Although, the answer score may not be a feature to be used alone, because there may still be relevant answers with low score, we hypothesize that such kind of social feature deserves to be investigated to improve retrieval performance.

In this paper, we  build upon CROKAGE~\cite{silva2019recommending,crokage2020}, and investigate  a novel approach  \textbf{CRAR} - \textbf{CR}owd  \textbf{A}nswer \textbf{R}ecommender that introduces other features and a novel way to integrate them. CROKAGE  first mines candidate answers from the crowd (i.e., Stack Overflow), whereas CRAR first mines relevant threads. This different approach enables the evaluation of  our hypothesis that more relevant threads in terms of popularity may contain more relevant answers. Besides, the information of all answers in the same thread would help the retrieval of relevant answers for the programming task because they could potentially bridge the semantic gap of the query and the answer, providing new terms that are not present in the chosen answer. Nonetheless, this hypothesis may not necessarily hold in all cases because there could be potentially negative answers that could also introduce some bias to associate the query with a relevant chosen answer, and thus the retrieval effectiveness could be impaired. Moreover, in this work, we investigate the effectiveness of several other features, such as, word embeddings, sentence embeddings (including a Convolution Neural Network), social features, and antonyms, which we motivate in Section~\ref{sec:extending}. 

In the end,  CRAR, which is the integrated approach with the best evaluated metrics  among those that we investigated using the above features, outperforms CROKAGE's results in terms of Top-10 accuracy, precision, recall, and reciprocal rank by 7\%, 6\%, 6\%, and nominally 0.11, respectively. Furthermore, we show how CRAR outperforms other baselines on the 
retrieval of relevant solutions for programming tasks. 

The main contributions of this work are as follows:

\begin{itemize}

\item A novel approach for integrating features from a large design space (CRAR) that recommends solutions for programming tasks expressed as queries, which outperforms the state-of-art CROKAGE~\cite{silva2019recommending,crokage2020}.

\item An empirical evaluation of a rich set of features, including semantic ones, such as word embeddings and sentence embeddings, including Convolution Neural Networks, to be combined in a relevance score for the recommendation of solutions for programming tasks. We show that the proposed combined score outperforms the individual ones, showing that they seem to work in a complementary way to rank the best answers. CRAR outperforms other baselines, including the state-of-art CROKAGE, in retrieving relevant answers for given programming tasks.

\item A replication package 
of CRAR, including the used dataset and the  instructions for  reproduction, so that other researchers can repeat and improve our results \cite{rodrigo_f_silva_2021_5115300}. 

\end{itemize}

The rest of this paper is structured as follows. Section~\ref{sec:extending} describes the current state-of-art work in retrieving relevant solutions for given programming tasks. Section~\ref{sec:architecture} shows the technical details of the proposed approach. Section~\ref{sec:experiments} shows the experimental design and the obtained results. Section~\ref{sec:discussion} provides a qualitative discussion of CRAR performance. Section~\ref{sec:threats} discusses the threats to validity of our experiments. Section~\ref{sec:relatedWord} presents the
related works. Finally, Section~\ref{sec:conclusion} concludes the paper and mention future works.

\section{State-of-the-Art Limitations}\label{sec:extending}





Given a programming task written in natural language, CROKAGE applies the BM25~\cite{robertson1994some} function to search for candidate Q\&A pairs over the index. Then, CROKAGE employs four factors to rank the most relevant answers from the pairs: semantic similarity, TF-IDF similarity, API method score and API class score. 
Crokage uses three API recommendation systems, BIKER~\cite{huang2018api}, NLP2API~\cite{rahman2018effective} and RACK~\cite{rahman2016rack}, to obtain a ranked list of relevant APIs of each programming task, and calculate the API class score for each candidate Q\&A pair. For API method, Crokage uses regular expression to obtain the used methods in each Q\&A pair, and identify the most frequent API methods for ranking Q\&A pairs.  



We have noticed important limitations on CROKAGE~\cite{silva2019recommending} approach. Relevant answers contained in more important discussions may be missed, while non relevant answers may be promoted because CROKAGE disregards some features from threads, such as their popularity,  as shown in Table~\ref{tab:soresultsquery}. However, popularity may correlate with age because older posts have more time to get voted. So, there may exist brand-new (non-popular) answers with high quality updated content, indicating that this kind of feature deserves investigation.

Other example of  feature disregarded by CROKAGE is the text of the whole thread. For instance, the query \textit{``How do I convert angle from radians to degrees?"} has five relevant answers from two threads\footnote{https://stackoverflow.com/questions/5763841}\footnote{https://stackoverflow.com/questions/9970281}. CROKAGE did not recommend any of those answers in the top-10. However, we observe that this query matches some words with the title and question body fields, suggesting an opportunity investigate the use of the whole thread text.

The inclusion of thread-related features could have improvements even in recall. The query \textit{``Generate random integer"} has 48 answers on goldset, and 28 of them are from the same thread\footnote{https://stackoverflow.com/questions/363681}. It is a popular thread with several answers with high score, but CROKAGE had recommended an answer from this thread only in the 98$^{th}$ position. So, the adoption of a feature based on the popularity of threads seems to deserve investigation.

 We also observe that lexical similarity tends to recommend some answers with an opposite semantic for the query intent. For example, in the query \textit{``fill array with filling char"}, CROKAGE included the thread title \textit{``how can I empty the unwanted position of the array?"} in their top-10. We observe that the terms ``\textit{empty}" and ``\textit{fill}" are antonymous. So, a feature to filter queries based on antonyms also seems to deserve investigation. 


\begin{figure}[]
\centerline{\includegraphics[width=0.90\textwidth]{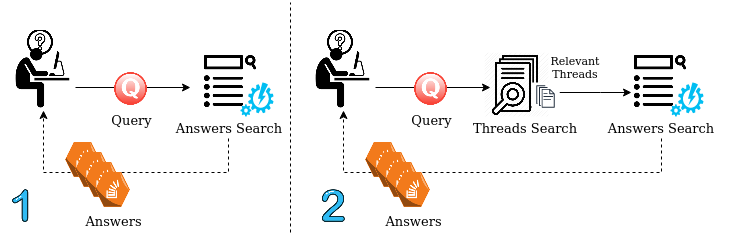}}
\caption{Overview of  CROKAGE (1) and CRAR (2). CROKAGE performs the retrieval over Stack Overflow answers. CRAR instead, first retrieves the relevant threads and then retrieves the answers containing in the relevant threads}
\label{fig:crokageXdycrokage}
\end{figure} 

In this work, we address the aforementioned limitations. Unlike CROKAGE~\cite{silva2019recommending}, CRAR considers the importance of the Stack Overflow Q\&A threads before selecting candidate answers as shown in Figure~\ref{fig:crokageXdycrokage}. So, given a programming task, the tool first retrieves the most relevant threads from a pre-built index, and then, over the retrieved threads, selects the relevant answers. CRAR employs a renewed variety of features to rank candidates threads and answers (described in Section~\ref{features}), which  mitigates the lack of term similarity between the query and candidate answers, and considers the presence of confounding antonyms and the importance of the discussions in which the answers are contained.



Thus, the present work focuses on the retrieval mechanism to provide the appropriate answers for programming tasks.

\section{CRAR Architecture} \label{sec:architecture}

CRAR is composed by three main phases, each composed by steps. The first two phases are responsible for the offline construction of several data structures. The third phase use those data structures to search for relevant answers. 

In the first phase (Section~\ref{buildAntonymsFromLists}), a dictionary of words and their antonyms are constructed by merging three repositories of antonyms. The second phase, described in Figure~\ref{fig:offline} and detailed in Section~\ref{offline},  uses data from Stack Overflow to construct models, maps and indices. The third phase, described in Figure~\ref{fig:online} and detailed in Section~\ref{online},  uses the generated data structures to search for relevant answers for programming tasks (i.e., queries). We describe the three phases as follows.

\subsection{Constructing the Dictionary of Antonyms} \label{buildAntonymsFromLists}

In this phase, an antonym dictionary is constructed containing a list of words and their respective antonyms, relying on  three external repositories. The first one\footnote{https://github.com/taikuukaits/SimpleWordlists} is an audited collection, obtained from a repository\footnote{https://github.com/airshipcloud/dictionary-seed/tree/master/wordnet/Thesaurus} which was built by the use of WordNet \cite{miller1995wordnet}. The second repository\footnote{https://gist.github.com/maxtruxa/b2ca551e42d3aead2b3d} is a list of common terms used in programming. The third repository is a sub product of a previous work~\cite{faruqui:2015:non-dist} on the construction of interpretable word vectors from hand-crafted linguistic resources. After downloading the three lists containing the words and their antonyms,  the lists are merged. The result is a dictionary containing the 14k words and their antonyms. This dictionary is later used to support one of our adopted features (further described in Section~\ref{features}).


\subsection{Constructing Models, Maps, and Indices} \label{offline}

Figure \ref{fig:offline} shows the construction of the models, maps and indices that are later used to support the search for relevant answers. The steps of the process occurs offline as follow:


\begin{figure}[]
\centerline{\includegraphics[width=0.9\textwidth]{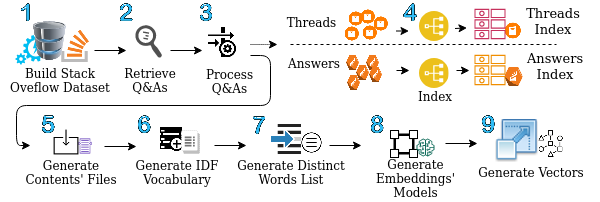}}
\caption{Offline process showing the construction of the models, maps and indices that are later used on the search for relevant answers.}
\label{fig:offline}
\end{figure}

\textbf{1) Build Stack Overflow Dataset}: The Stack Overflow data  dump (march 2019) \footnote{https://archive.org/details/stackexchange - dump published in March 2019} is downloaded and a table with all posts (questions and answers) are loaded into a local PostgreSQL database. 

\textbf{2) Retrieve Q\&As}: We employ regular expressions to analyse the “tags”. For Java language, we select all the questions containing “java” but not “javascript”. Thus,  all posts related to Java are retrieved, resulting in a total of 1,543,653 questions and 2,562,484 answers. 

\textbf{3) Process Q\&As}: For each loaded post,  the code is separated from the texts using the tags \texttt{<code>} and \texttt{<pre>}, and perform  simple natural language processing steps over the texts. All punctuation symbols, stop words\footnote{https://bit.ly/1Nt4eMh}, small words (i.e., length lower than two), numbers and extra spaces are removed. The processed text is saved alongside the original text of each question and answer, so that later we use their original version to recommend relevant answers. The output of this step is the two sets (i.e., whole threads and answers only).
 
\textbf{4) Index}: After processing all questions and answers, two indices are built: one for threads and one for answers. To build the threads' index, all questions are selected in such a way that they contain answers and score higher than zero, discarding the others. \red{Because, the index is based on bag-of-words, the order in which answers are processed is not relevant.}  Then, for each selected question, its answers containing code examples and score higher than zero are selected, and the other answers are discarded. A thread is reconstructed containing its question's pre-processed information and its pre-processed bodies of the selected answers (i.e., text + code examples). Finally, three social features are included: the question score, the number of selected answers and the sum of all answers' scores. The processed threads are stored in the Lucene~\cite{Lucene} index. This static index will be used as a starting point to retrieve the most relevant threads for a programming task.  

To build the answers' index, answers are selected in such a way that they contain code examples, score higher than zero and a parent question with their score. A document for each answer containing their pre-processed information (i.e., body text + code examples) along with the information of their parent question (i.e., title + body text + code examples) is added to a Lucene~\cite{Lucene} index. This static index will be used to retrieve the most relevant answers for a given query.  

\textbf{5) Generate Contents' Files}: In this step, given the pre-processed threads from Step 3, two text files are generated. The first file contains the titles of the questions, one title in each line. The second file contains one thread per line. The content of each thread includes title, body text, and code examples of both question and answers, respectively.  

\textbf{6) Generate IDF Vocabulary} and \textbf{7) Generate the Distinct Words List}: Inverse Document Frequency (IDF) means the inverse number of documents containing a word. The metric has been used to measure the importance of a word in a corpus~\cite{xu2017answerbot,silva2019recommending,huang2018api}, i.e., infrequent words across the corpus carry more importance than more frequent words. In Step 6, the IDF value  is calculated from the second file generated in the previous step containing the pre-processed information of all threads, and a map containing each word and its IDF value is constructed. This map will be used later  (Section~\ref{features}) to calculate the asymmetric similarity. Step 7 builds  a list containing the 1,176,362 non-repeated words of the vocabulary, which will be used later to generate  word vectors (Step 9).



\textbf{8) Generate Embeddings' Models}: The keywords of the programming task (i.e., query) and the answers from Stack Overflow might not always match. To overcome this lexical gap issue we evaluate two kind of models capable of capturing the semantics of words and sentences.
\\
\textbf{A. Word Embeddings}. We use \textit{FastText}~\cite{bojanowski2016enriching}, a widely adopted~\cite{silva2019recommending,efstathiou2019semantic,rahman2018effective,sachdev2018retrieval} word embedding technique in software engineering. We employ \textit{FastText} over the contents file (generated in Step 5) to construct a skip-gram model with the following parameters: vector size=100 (as recommended in Fasttext tutorial\footnote{https://fasttext.cc/docs/en/unsupervised-tutorial.html}), epoch = 20 (higher than the default of five epochs in Fasttext tutorial to ensure possibly more effectiveness), minimum size = 2, maximum size = 5 (empirically improved result, with one unit less than the tutorial, which mentioned that other languages could have other values), and finally, left the other parameters with  default values. We adopt the skip-gram model over the CBOW model because it has been observed to be more efficient with subword information~\cite{bojanowski2016enriching}. Like CROKAGE~\cite{silva2019recommending}, words are not stemmed, since \textit{FastText} looks for subwords.
\\\textbf{B. Sentence Embeddings}. In this case, a full sentence is embedded into a vector space. The motivation for sentence embedding is to capture semantic similarity between two sentences that would go beyond individual words. For this kind of embedding, we evaluate two solutions:  \textit{Sent2Vec}~\cite{pagliardini2017unsupervised}, which is an extension of \textit{FastText} that can learn the embeddings for n-grams, and an adapted \textit{CNN} for sentence embedding~\cite{liu2019learning}. 
\\
\textbf{B1. Sentence Embedding - Sent2Vec}.
Unlike \textit{FastText}, \textit{Sent2Vec} considers the order of the words. We customize the model using the same parameters as for \textit{FastText} and set the n-grams parameter to three. The output of this step is the  model used next to generate words' representations in the form of high-dimensional vectors. 
\\
\textbf{B2. Sentence Embedding - CNN}. \textit{Convolutional Neural Networks (CNN)} is a class of Deep Neural Networks. We use LeNet5~\cite{lecun1998} \textit{CNN} implementation to identify similar words, using unsupervised learning. The input is two-dimensional numeric vectors. One dimension is the embedding of each word, using similar parameters as for \textit{FastText}. The other dimension is the number of words, with token number equals 10. The output is a sentence vector corresponding to the input words. The training phase produces a vector space of all questions titles. 
This \textit{CNN} uses as parameters: epoch = 20, learning rate = 1e-2, nodes in hidden layers = 1000. These parameters were the same used by Liu et al.~\cite{liu2019learning} to embed method bodies into a vector space.


\textbf{9) Generate Word Vectors}: this  step uses the models constructed in the previous step to represent words and sentences by high-dimensional vectors. First, the vector representation for each word of our vocabulary is generated by \textit{FastText} function called \textit{print-word-vectors} passing as parameters the list of the distinct words of our vocabulary (generated in Step 7) and the skip-gram model (generated in the previous Step). 
The resulting map (word - embeddings) is used later, together with the IDF map (generated in Step 6), to calculate the asymmetric similarity between two given bag-of-words. Second, similarly to \textit{FastText}, the vector representation for all questions' titles is generated using the function of \textit{Sent2Vec} called \textit{print-sentence-vectors} passing the \textit{Sent2Vec} model produced in the previous step and a file containing the titles of all questions. Likewise, a similar map is generated for all questions' titles and their respective embeddings generated by the CNN.
These maps are used later to calculate the sentence similarity between a query and a question title.

Whenever, we need to update the Stack Overflow dump, we should re-run this process shown in Figure~\ref{fig:offline}, which should take around one processing day.

\subsection{Searching for Relevant Answers} \label{online}
This \textit{online} phase use the \textit{offline} data generated in the two previous phases.
Figure \ref{fig:online} shows the search for relevant answers. We describe each of its step as follows:

 \begin{figure}[t]
\centerline{\includegraphics[width=0.95\textwidth]{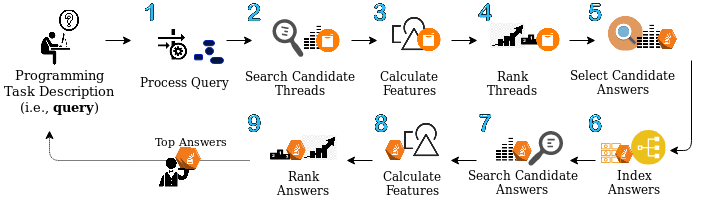}}
\caption{Searching for relevant answers. The search starts with the programming task provided by the user. In the end of pipeline, the approach returns the top N most relevant answers from Stack Overflow for the task, where N is an input parameter}
\label{fig:online}
\end{figure}

\textbf{1) Process Query}: The programming task provided by the user can contain words that are not helpful for the search for relevant answers. In this step,  those terms are removed. Standard natural language processing are performed on the query, such as removing all punctuation symbols, stop words, one-character\footnote{One/two-character word removal has worsened the overall result, explaining our decision for one-character words.}  words, which are words that usually repeat in many different tasks, and numbers. The result is a non-repeated  bag-of-words representing the task.


\textbf{2) Search Candidates Threads}: In Step 2, given the bag-of-words $Q$ representing the query and the threads' index (Section~\ref{offline}), CRAR uses BM25~\cite{robertson1994some} function to determine the lexical similarity between each thread $T$ from the index and $Q$ as follows:

\vspace{-0.7em}
{\small
\begin{equation}
\begin{aligned}
lexicalSim(T,Q)={} & \sum_{i=1}^{n} idf(q_i) \\ \ast
                   & \frac{f(q_i,T)  * (k+1)}{f(q_i,T)+k*(1-b+ b*\frac{|T|}{avgdl})}
\end{aligned}
\label{eq:bm25}  
\end{equation}
}
\vspace{-0.3em}

 \noindent where $|T|$ is the length of the thread $T$ in words, $f(q_i,T)$ is keyword $q_i$'s term frequency in thread $T$ and $avgdl$ is the average document length in the index. $b$ and $k$ are two parameters where $k$ is a non-negative finite value that controls non-linear term frequency normalization (saturation), and $b$ controls to what degree document length normalizes term frequency values, whose range is in $[0,1]$. CRAR customizes these parameters: $k=1.2$ and $b=0.9$. The function $idf(q_i)$ is the inverse document frequency of keyword $q_i$.
 



\textbf{3) and 8) Calculate Features}: After obtaining the candidates (i.e., threads or answers), CRAR determines the similarity between the query and each candidate by using several features. We describe all the evaluated features in Section~\ref{features} and the set of features used in each step 3 or 8 in Section~\ref{baselines}. Step 3 uses part of the features to rank the whole thread, independently of which answer(s) contained in the thread would be selected to be part of final answer. Step 8 uses other part of the features to rank only \textit{answers} that will be used as final result.  CRAR determines the scores for each feature, normalize this score and then calculates a final combined score as a weighted combination of all features' scores, as follows:

\vspace{-0.5em}
\begin{equation}
finalScore(C,Q) = \sum_{i}^{n} s_{f_i} * w_{f_i}
\label{eq:finalScore}
\end{equation}
\vspace{-0.5em}

\noindent where $s_{f_i}$ denotes the score of the feature $f_i$ and $w_{f_i}$ denotes the weight associated with this feature score.

\textbf{4) and 9) Rank Candidates}: given a list of candidates (threads (Step 3) or answers (Step 8)), CRAR sorts them according to their final score (Formula~\ref{eq:finalScore}) in descending order. 




\textbf{5) Select Candidate Answers} and \textbf{6) Index Answers}: Given a ranked list of candidate threads, CRAR retrieves only their answers (Step 5) having score higher than zero and contain code examples. Next, in Step 6, CRAR constructs a Lucene Index~\cite{Lucene} containing  this set of answers, the same way as in Section~\ref{offline}.

\textbf{7) Search Candidate Answers}: Given an answers' index built online (Step 6), CRAR uses the function $lexicalSim(A,Q)$ (Formula~\ref{eq:bm25}) to calculate the lexical similarity between each answer $A$ from the index and the bag-of-words $Q$ representing the query. The function retrieves the top K answers sorted by their relevance to the query, which is the output of this Step.  We determine the threshold value of K as 150, since we are considering only answers, and higher values for K  in general lead to very dissimilar answers.

In the end (Step 9), given an input parameter N, CRAR selects the top N answers and returns them as the final output.

\subsection{Adopted Features}\label{features}
Given the bag-of-words \textit{Q} representing the query - Section~\ref{online} - Step 1) and a set of candidates (threads or answers), CRAR uses several features to calculate the similarity between each candidate against \textit{Q}. This similarity is calculated using the candidate's content according to the feature. If the candidate is a thread (as in Figure~\ref{fig:online}, Step 3), we represent each thread's content as the bag-of-words \textit{T}. If the candidate is an answer (as in Figure~\ref{fig:online}, Step 8), we represent each answer as the bag-of-words \textit{A}. CRAR calculates a score between two documents \textit{Q} and (\textit{T} or \textit{A}) for each of the following features:


\textbf{Antonyms}: This feature penalizes the candidate (question or answer) containing the antonyms of the query. For instance, if the query is about \textit{``how to uncompress a file"}, questions/answers containing ``compress" are penalized. CRAR calculates the feature as follows. First, CRAR employs Stanford Part-Of-Speech Tagger (POS Tagger)~\cite{stanfordnlp} to annotate each word of \textit{Q}. Then, it uses the dictionary of antonyms (Section~\ref{buildAntonymsFromLists}) to collect the antonyms of the nouns and verbs of \textit{Q}. If the query contains antonyms in itself (e.g., ``How to \textit{zip} / \textit{unzip} a file in Java?"), CRAR sets to this feature a zero score. Otherwise, it merges the collected antonyms into a list of non-repeated bag-of-words $A_{nt}$. Next, CRAR constructs a bag-of-words $C_{an}$ using the pre-processed information of the candidate. For threads, CRAR concatenates the title (i.e., question title) and the body text (i.e., question body text). For answers, it concatenates the title of the parent question, the body text and the code. After constructing $A_{nt}$ and $C_{an}$, CRAR computes the antonyms score as follows: 

\vspace{-0.6em}
\begin{equation}
AntonymsScore(A_{nt},C_{an}) = |A_{nt} \cap C_{an}|
\label{eq:AntonymsScore}
\end{equation}

\noindent



\textbf{TF:} TF stands for term frequency (TF) and represents the number of times a term (i.e., word) occurs in a document. This feature is used to calculate the lexical similarity between the query and each candidate thread based on their common shared words. 
CRAR determines the TF Score between the two documents \textit{Q} (query) and \textit{T} (question title, question text body,  answers' text body and  answers' code) as follows: 

\vspace{-0.6em}
\begin{equation}
tfScore(Q,T) {}= \frac{d_Q \cdot d_T}{|d_Q| \cdot |d_T|} 
= \frac{\sum_{1}^{N} tf_{tu,dq} \cdot tf_{tu,dt}}{\sqrt{\sum_{1}^{N} tf_{ti,dq}^{2}} \cdot \sqrt{tf_{ti,dt}^{2}}}
\label{eq:tfscore}
\end{equation}
\vspace{-1em}

\noindent where $d_Q$ denotes the query and $d_T$ represents the thread. $tf_{tu,dk}$ represents the number of times each term $t~\epsilon~U$ appears in document \textit{dk} and $tf_{ti,dk}$ denotes the number of times each the term $t$ appears in document $dk$.

\textbf{TF-IDF:} TF-IDF stands for the product of the TF and the IDF. We use this feature to determine the lexical similarity between the query and each candidate answer. 

CRAR calculates the TF-IDF for each word $W$ of a document (query or answer) as follows:

\vspace{-0.6em}
\begin{equation}
\mathit{TF\mbox{-}IDF(W)} = TF(W) * log_{10}(\frac{N}{df_{w}})
\label{eq:idf}
\end{equation}
\vspace{-0.3em}

\noindent where $N$ denotes the total number of documents in the corpus, and $df_{W}$ denotes the number of documents containing the word $W$. CRAR determines the lexical similarity between $Q$ and the document $A$ using their cosine similarity as follows:

\vspace{-1em}
\begin{equation}
\begin{aligned}
tfidfScore(Q,A) {}= & \frac{d_Q \cdot d_A}{|d_Q| \cdot |d_A|} \\
= & \frac{\sum_{1}^{N} tfidf_{ti,dq} \cdot tfidf_{ti,da}}{\sqrt{\sum_{1}^{N} tfidf_{ti,dq}^{2}} \cdot \sqrt{tfidf_{ti,da}^{2}}}
\label{eq:idfscore}
\end{aligned}
\end{equation}

\noindent where $d_Q$ refers to the query, $d_A$ represents the candidate answer and $tfidf_{ti,dk}$ is the term weight for each word of the document \textit{dk}.

\textbf{Sentence Similarity with \textit{Sent2Vec} or \textit{CNN}:} CRAR determines a semantic similarity score between \textit{Q} (i.e., the query) and each Q\&A pair by comparing two sentences: \textit{Q} and the title \textit{T} of the pair. For this, CRAR uses the cosine distance between their embeddings' vectors, which are generated by \textit{Sent2Vec} or \textit{CNN} embeddings as follows: 

\vspace{-0.6em}
\begin{equation}
\begin{aligned}
sentScore(Q,T)= & \frac{Vec1 \cdot  Vec2}{|Vec1|\times |Vec2|} \\ 
               = & \frac{\sum_{1}^{n}Vec1_i \times Vec2_i}{\sqrt{\sum_{1}^{n}Vec1_i^{2}} \times {\sqrt{\sum_{1}^{n}Vec2_i^{2}}}}
\label{eq:semscore}
\end{aligned}
\end{equation}
\vspace{-0.6em}

\noindent where $VecK$ is the 100-dimensional vector representing the sentences $Q$ or $T$ and $VecK_i$ is the element of the vector $VecK$ at position $i$.

\textbf{Asymmetric Similarity:} CRAR determines another semantic similarity score between \textit{Q} and each candidate (thread or answer). For the threads, CRAR builds two bag-of-words $T_1$ and $T_2$ where $T_1$ is composed by the thread' title, while $T_2$ is composed by the concatenation of the thread' body and its answers' body. For the answers, CRAR builds a bag-of-words \textit{A} composed by the answer body and the title of its parent question. CRAR then determines the asymmetric similarity relevance between \textit{Q} and a bag-of-words \textit{T}, where $T~\epsilon~(T_1,T_2,A)$ as follows:

\vspace{-0.6em}
\begin{equation}
    asym(Q \rightarrow T) = \frac{  \sum_{ w \in Q }{sim(w,T)} * idf(w) }{\sum_{ w \in Q }{idf(w)} }
\label{eq:asymmetricsim}    
\end{equation}
\vspace{-0.6em}

\noindent where $sim(w,T)$ is the maximum value of $sim(w,w_T)$ for every word $w_T \in T$. $sim(w,w_T)$ is the cosine similarity between $w$ and $w_T$ embedding vectors and $idf(w)$ is the correspondent IDF value of the word $w$. The other asymmetric relevance namely $asym(T \rightarrow Q$) can be calculated by swapping $Q$ and $T$ in Formula~\ref{eq:asymmetricsim}. Thus, the final similarity between the task $T$ and $Q$ is the harmonic mean of the two asymmetric relevance scores as follows: 


\vspace{-0.3em}
\begin{equation}
    asymScore(Q,T) = \frac{2*asym(Q\rightarrow T)*asym(T\rightarrow Q))}{asym(Q\rightarrow T)+asym(T\rightarrow Q)}
\label{eq:asymmetricsimfinal}    
\end{equation}


\textbf{Top Method:} If an API method appears across multiple candidates answers, it is more likely to contain a proper solution for the query problem~\cite{silva2019recommending,huang2018api}. CRAR rewards candidate answers containing the most common method in their answers. For this, CRAR obtains the methods of each candidate answer using regular expressions and assigns in each answer a method score as follows:

\begin{equation}
methodScore(A) = \frac{log_{2}(f_m)}{S}
\label{eq:methodScore}
\end{equation}

 \noindent where $f_m$ is the top method frequency and $S$ controls the scale of the score. We empirically test different values for $S$ and find that $S=10$ gives the best performance. If the answer does not contain the top method, the score is set to zero.



\textbf{Social Features:} social features had already been used by previous works~\cite{ponzanelli2014mining,xu2017answerbot} as factors to determine the relevance of Stack Overflow posts (question or answer). 
In Table~\ref{tab:soresultsquery}, we have shown a motivating example on why social feature could play a role in improving retrieval performance. Although, social features may exclude relevant content because there may still be recent relevant answers that could not have had sufficient time to get the score increased. 

We investigate the use three social features to identify relevant threads among the candidates: Answer Count (i.e., the number of answers), Question Score (i.e., upvotes minus downvotes), and Total Answer Score, which is the sum of all answers scores (i.e., upvotes minus downvotes). CRAR determines a value in range of [0,1] for each social feature.  Answer Count and Total Answer Score are normalized values between [0,1]. However, instead of the calculating  the Question Score with simple normalization, we opted for using degrees based on the Stack Overflow privileges\footnote{https://stackoverflow.com/help/privileges}, which awards privileges according a range on the number of points. CRAR uses a similar range to assign the value in [0,1] according to the respective range the score belongs, where the maximum range is 500+, which corresponds to score 1. The defined Question Score ranges and their respective values are shown in Table~\ref{tab:privileges}.

\begin{table}[t]
\centering
\caption{Question Score ranges with their respective values}
\label{tab:privileges}
\begin{tabular}{lcccc}
\hline
\textbf{Question Score range} & \textbf{Value} \\ \hline
$\leq$ 1              & 0.1                              \\
2-5              & 0.2                                  \\
6-10            & 0.3                             \\
11-25            & 0.4                             \\
26-50            & 0.5                             \\
51-75            & 0.6                             \\
76-100            & 0.7                             \\
101-200            & 0.8                             \\
201-500            & 0.9                             \\
$>$ 500            & 1.0                             \\
                      \hline
\end{tabular}
\end{table}

\section{Experimental Evaluation} \label{sec:experiments}


Herein, we evaluate our novel approach CRAR that given a programming task, recommends Stack Overflow answers containing code and explanations. We compare our approach against multiple other baselines (including the state-of-art CROKAGE~\cite{silva2019recommending}) where each baseline is constructed as a variation of our proposed set of features. However, since we adopted multiple features and their combination would result in a high number of baselines, we focus on evaluating features not explored by the CROKAGE. For the evaluation, we adopt their  ground truth\footnote{https://github.com/muldon/CROKAGE-replication-package} (i.e., a list containing 115 programming tasks and the relevant Stack Overflow answers for each task), as well as the same four performance metrics (i.e., Top-K Accuracy (Hit@K), Mean Reciprocal Rank (MRR@K), Mean Average Precision (MAP@K), Mean Recall (MR@K)), which have also been adopted by other previous works in the software engineering~\cite{rahman2019supporting,rahman2016rack, rahman2018effective,rahman2017strict,huang2018api,xu2017answerbot}. 
Those 115 Java programming tasks were selected from three Java tutorial sites: Java2s~\cite{java2s}, BeginnersBook~\cite{BeginnersBook} and KodeJava~\cite{KodeJava}. Queries based on those tasks were used as input to three search engines: Google~\cite{Google}, Bing~\cite{Bing} and Stack Overflow search~\cite{SOSearch}. Google and Bing queries were narrowed to \textit{site:stackoverflow.com}. Results from Stack Overflow internal search engine should pass a threshold of a minimum of 100 views.
For each of the 115 queries, results were merged and duplicates removed. Questions with no answers were discarded, and we selected answers with score highter than zero and containing code examples. This automatic process resulted into 6,558 answers. Then, two professional developers manually evaluated all these 6,558 answers. The final ground-truth contained 1,743 relevant answers for the 115 queries. 


Thus, we run all baselines to recommended the answers for the testing set. Then, we contrast their recommended answers against the ground truth and calculate the performance metrics (i.e., Top-K Accuracy, Mean Reciprocal Rank, Mean Average Precision and Mean Recall). In particular, we answer to the following research questions in our experiments:

\textbf{RQ1:} To what extent do Antonyms and Social Features influence the ranking of candidate answers?

\textbf{RQ2:} How the individual thread features can influence the ranking of candidate answers compared to the features combined (CRAR)?

\textbf{RQ3:} How the individual answer features can influence the ranking of candidate answers compared to the features combined (CRAR)?

\textbf{RQ4:} How does CRAR perform compared to the state-of-art CROKAGE in retrieving relevant answers for given programming tasks?

\subsection{Baselines and Results} \label{baselines}

We customize a baseline called \textit{Template} that represents the basic version of our approach as described in Section~\ref{online}. That is, given a programming task, the \textit{Template} baseline first searches for relevant threads and then, for relevant answers. \textit{Template} calculates features for threads and answers (Steps 3 and 8 of Figure \ref{fig:online} respectively) to determine their relevance against a query. This baseline calculates seven features' scores for threads and four features's scores for answers, as in Table \ref{tab:featuresxstage}. We describe how \textit{Template} calculates the score of each feature for the threads and answers as follows:


\textbf{Threads}: Given a set of N threads (i.e., N=500) as result of BM25~\cite{robertson1994some} search (i.e., Section~\ref{online} - Step 2 - Figure \ref{fig:online}), \textit{Template} calculates the features for the threads in two stages. In the first stage, \textit{Template} calculates the non-social features: the sentence similarity herein represented by Sent2Vec, the two asymmetric similarities and the TF. Then, \textit{Template} sums their scores (i.e., Formula \ref{eq:finalScore}) and use the resulting score to sort the candidate threads in descending order. \textit{Template} then selects the top 250 threads with the highest resulting score. In the second stage, \textit{Template} calculates the scores of all the seven thread's features over the remaining threads (i.e., 250) and determines the final score (i.e., Formula \ref{eq:finalScore}) for each candidate thread. \textit{Template} then selects the top 100 threads according to their final score and discards the others. \textit{Template} uses the same weight (i.e., 0.5) for the seven thread features (as in Table \ref{tab:featuresxstage}). 

\textbf{Answers}: Given a set of N answers (i.e., N$<=$150) as result of the selection of candidate answers (i.e., Section~\ref{online} - Step 7 - Figure \ref{fig:online}), \textit{Template} calculates the score for four features as in Table~\ref{tab:featuresxstage}: Asymmetric Similarity, TF-IDF, Top Method and the Thread Score as described above. \textit{Template} then uses the final score (i.e., Formula \ref{eq:finalScore}) to rank the top K answers (where K is a parameter) and delivers them as the final output (i.e., Section~\ref{online} - Step 9).



\begin{table}[]
\centering
\caption{Features used by the template baseline (i.e., \textit{Template}) to calculate the similarity between a thread or an answer against the programming task and their weights. }
\label{tab:featuresxstage}
\begin{tabular}{lccc}
\hline
\multicolumn{1}{c}{}                         & Feature              & Information                                                                                                    & W. \\ \hline
\multirow{7}{*}{Threads}                     &Sent2Vec/CNN  & Title                                                                                                          & 0.5    \\ \cline{2-4}
                                             & Asymmetric Similarity & Title                                                                                                          & 0.5    \\ \cline{2-4}
                                             & Asymmetric Similarity & Body + Answers’ Body                                                                                           & 0.5    \\ \cline{2-4}
                                             & TF                   & \begin{tabular}[c]{@{}c@{}}Title + Body + \\ Answers’ Body + \\ Answers’ Code\end{tabular}                     & 0.5    \\ \cline{2-4}
                                             & Social Feature       & \begin{tabular}[c]{@{}c@{}}Answer Count \\ (the number of answers)\end{tabular}                                & 0.5    \\ \cline{2-4}
                                             & Social Feature       & \begin{tabular}[c]{@{}c@{}}Total Answer Score\\  (the sum of answers scores) \\ 
                                             \end{tabular}                  & 0.5    \\ \cline{2-4}
                                             & Social Feature       & \begin{tabular}[c]{@{}c@{}}Question Score (the score \\ of the question)\end{tabular}           & 0.5    \\ \hline
\multicolumn{1}{c}{\multirow{4}{*}{Answers}} & Asymmetric Similarity  & \begin{tabular}[c]{@{}c@{}}Title of parent question +\\  Body\end{tabular}                                     & 1.0    \\ \cline{2-4}
\multicolumn{1}{c}{}                         & TF-IDF               & \begin{tabular}[c]{@{}c@{}}Title of parent question +\\  Body of parent question +\\  Body + Code\end{tabular} & 0.5    \\ \cline{2-4}
\multicolumn{1}{c}{}                         & Top Method           &                                                                                                                & 0.75   \\  \cline{2-4}
\multicolumn{1}{c}{}                         & Thread Score         &                                                                                                                & 0.75   \\ \hline
\end{tabular}
\end{table}

We justify our choice for using TF to select threads and TF-IDF to select answers due to  preliminary tests on the retrieval performance. We empirically tested the combinations of the two functions in the two stages and showed only the best combination for those functions, for the sake of presentation. Likewise, the weights associated to each feature and the number of threads filtered in each stage as described above were also empirically tested. 

After constructing \textit{Template}, we then extend it to generate other baselines by adding/removing features as follows: 

\textbf{Template-Without-SF}: herein, we disable the Social Features (SF) from the list of thread's features (Table \ref{tab:featuresxstage}), thus relying only on the other four non-social features during the features calculation (Section~\ref{online} - Step 3).

\textbf{Template-SF-SocialFeatureType}: we add the combination of Social Features and construct six baselines. The main reason is to verify how each combination impacts the retrieval performance. The feature types are: 

\begin{itemize}
\item SocialFeatureType: \textit{TAS} for Total Answer Score, \textit{QS} for Question Score and \textit{AC} for Answers Count. We analyze all possible combinations of these features. For example, \textit{TAS-QS} are \textit{Total Answer Score + Question Score}, discarding the Social Feature \textit{Answer Count}. The combination of these three features are represented on \textit{Template} baseline, i.e. its the same of \textit{Template-TAS-AC-SF}.
\end{itemize}



\textbf{Template-Ant-POS-CandidateType}: we add the antonyms (Ant) feature to \textit{Template} and construct nine baselines. We use this feature as a filter and discards threads and/or answers containing antonyms for the query (i.e., the $AntonymsScore > 0$ - Formula~\ref{eq:AntonymsScore} - Section~\ref{features}). The baselines are the combination of the Part of Speech used to generate antonyms and the candidate type as follows: 

\begin{itemize}
\item Part of Speech: NN for nouns, VB for verbs or NN\_VB for both.
\item Candidate type: threads (TR), answers (ANS) or both (TR\_ANS).
\end{itemize}

For example, the baseline \textit{Template-Ant-NN\_VB-TR\_ANS} discards threads (TR) and answers (ANS) containing antonyms for the query where the antonyms are calculated for the nouns (NN) and the verbs (VB) of the query. 


\textbf{RQ1:} \textit{To what extent do Antonyms and Social Features influence the ranking of candidate answers?}

We run each baseline against the testing set queries to search for relevant answers (Section \ref{online}) on the 2019 march Stack Overflow dataset, as described in Section~\ref{offline}) and collect the metrics for Social Features as shown in Table~\ref{tab:socialfeatures} and for Antonyms in Table~\ref{tab:anthonyms}. We sort the baselines by the sum of the four metrics. 

\begin{table}[t]
\centering
\caption{Performance of Social Features combined over 7 baselines in terms of Hit@K, MRR@K, MAP@K, and MR@K, for K=10}
\label{tab:socialfeatures}
\begin{tabular}{lcccc}
\hline
\textbf{Baseline Approach} & \textbf{Hit} & \textbf{MRR} & \textbf{MAP} & \multicolumn{1}{c}{\textbf{MR}} \\ \hline
Template-Without-SF  & 0.75 & 0.44 & 0.39 & 0.20 \\
Template-SF-TAS      & 0.81 & 0.52 & 0.46 & 0.23 \\
Template-SF-TAS-AC   & 0.84 & 0.55 & 0.48 & 0.25 \\
Template-SF-QS       & 0.86 & 0.54 & 0.48 & 0.23 \\
Template-SF-AC       & 0.88 & 0.53 & 0.47 & 0.23 \\
Template-SF-QS-TAS   & 0.88 & 0.56 & 0.48 & 0.25 \\
Template (SF-All)    & 0.88 & 0.60 & 0.50 & 0.24 \\ 
Tempĺate-SF-QS-AC    & 0.89 & 0.61 & 0.51 & 0.26 \\
                      \hline
\end{tabular}
\end{table}

\begin{table}[t]
\centering
\caption{Performance of Anthonyms over 9 baselines in terms of Hit@K, MRR@K, MAP@K, and MR@K, for K=10}
\label{tab:anthonyms}
\begin{tabular}{lcccc}
\hline
\textbf{Baseline Approach} & \textbf{Hit} & \textbf{MRR} & \textbf{MAP} & \multicolumn{1}{c}{\textbf{MR}} \\ \hline
Template-Ant-NN\_VB-TR\_ANS     & 0.84         & 0.54         & 0.48        & 0.22 \\
Template-Ant-VB-ANS             & 0.86         & 0.56         & 0.50        & 0.21 \\
Template-Ant-NN-TR\_ANS         & 0.86         & 0.57         & 0.50        & 0.21 \\
Template-Ant-NN\_VB-TR          & 0.88         & 0.57         & 0.51        & 0.25 \\
Template-Ant-VB-TR\_ANS         & 0.88         & 0.58         & 0.50        & 0.26 \\   
Template-Ant-NN\_VB-ANS         & 0.88         & 0.58         & 0.52        & 0.23 \\
Template                        & 0.88         & 0.60         & 0.50        & 0.24 \\
Template-Ant-VB-TR              & 0.88         & 0.61         & 0.53        & 0.23 \\
Template-Ant-NN-TR              & 0.89         & 0.58         & 0.50        & 0.26 \\  
Template-Ant-NN-ANS             & 0.89         & 0.62         & 0.53        & 0.26 \\   
                      \hline
\end{tabular}
\end{table}

\textit{Social Features}: as shown in Table~\ref{tab:socialfeatures}, the presence of the Social Features strongly influences the ranking of candidate answers in our baselines. The difference between the version without the social features (i.e., \textit{Template-WithoutSF}) and the version containing all of them (i.e., \textit{Template}) in terms of Top-K Accuracy, Mean Average Precision, Mean Recall, and Mean Reciprocal Rank for K=10 is by 13\%, 11\%, 4\% and 0.16 respectively (in absolute values). This influence is statistically attested~\cite{wilcoxon1945individual} (i.e., p-values $<$ 0.05) for all metrics with effect size ranging from 0.31 (medium) to 0.53 (large), calculated with $r=Z/\sqrt{n}$ \cite{fritz2012effect}. These results suggest that social factors (i.e., features) help the retrieval of relevant answers by promoting important threads. 

\textit{Antonyms}: The baseline Template-Ant-NN-ANS has the best overall performance for antonyms as shown in Table~\ref{tab:anthonyms}. \textit{Template-Ant-NN-ANS} is the baseline  with lexical, semantic and social features (\textit{Template}), combined with the antonym feature (\textit{Ant}) that discard  answers (\textit{ANS}) containing antonyms for the nouns (\textit{NN}) in the query. Although most of the baselines containing antonyms perform worse than the template baseline (i.e., \textit{Template}), little gain can be obtained with antonyms (i.e., up to 1\% in Top-K Accuracy, 3\% in Mean Average Precision, 2\% in Mean Recall, and a Mean Reciprocal Rank of 0.02). No significant difference (i.e., p-values $<$ 0.05) using non-parametric Wilcoxon signed-rank test~\cite{wilcoxon1945individual} is observed for any metric between \textit{Template} and \textit{Template-Ant-NN-ANS} suggesting that the use of antonyms does not improve the performance in the ranking of candidate answers significantly. 

The Table~\ref{tab:performance} shows the performance of CRAR baseline compared with two other baselines, the \textit{Template} baseline, and the \textit{Template-Ant-NN-ANS$|$SF-QS-AC} which contains Anthonyms (\textit{Answers + Nouns}) and a combination  of \textit{Question Score + Answer Count}. We observe that  \textit{Template-Ant-NN-ANS$|$SF-QS-AC} and \textit{Template-Ant-NN-ANS$|$SF-All} are basically tied because, the former has better \textit{Hit} and \textit{MR}, and the latter has better \textit{MRR} and \textit{MAP}. We arbitrarily opted to choose CRAR as the combination with all social features, and use it in the following analyses.

\begin{table}[t]
\centering
\caption{Performance of best Anthonyms and Social Factors combination in terms of Hit@K, MRR@K, MAP@K, and MR@K, for K=10}
\label{tab:performance}
\begin{tabular}{lcccc}
\hline
\textbf{Baseline Approach} & \textbf{Hit} & \textbf{MRR} & \textbf{MAP} & \multicolumn{1}{c}{\textbf{MR}} \\ \hline
Template (SF-All)                     & 0.88     & 0.60    & 0.50    & 0.24 \\
         
Template-Ant-NN-ANS$|$SF-QS-AC  & 0.91     & 0.61    & 0.52    & 0.27 \\   
Template-Ant-NN-ANS$|$SF-All (CRAR) & 0.89     & 0.62    & 0.53    & 0.26 \\ 
\hline
\end{tabular}
\end{table}



\vspace{.5em}
\begin{center}
\fbox{\begin{minipage}{32em}
\textbf{RQ1 Answer:} Antonyms and Social Features can improve the ranking of candidate answers. However, the gains found are not significant for Antonyms but are positively strong for the use of Social Features which significant gains.
\end{minipage}}
\end{center}
\vspace{.5em}

\textbf{RQ2:} \textit{How the individual thread features can influence the ranking of candidate answers compared to the features combined (CRAR)?}

We investigate how each of the four individual thread non-social features (i.e., Table~\ref{tab:featuresxstage}) influences the ranking of candidate answers by comparing the individual performance of the features against their combination (CRAR). For this, we construct 11 baselines as follows: 

\noindent \textbf{Baselines:} We construct four baselines representing each individual non-social thread feature of \textit{Template-Without-SF}, where in each baseline we keep the weight of its representing feature and set the weights of the other features to zero. For example, for the baseline \textit{Asym. Sim. - Title}, we set the weights of all thread features to zero, except for \textit{Asym. Sim. - Title} feature. We construct other four baselines, each of them named \textit{CRAR Without $<$Feature(s)$>$} to measure the CRAR performance without each respective feature. For example, for the baseline \textit{CRAR Without Asym. Sim. - Title}, we include all thread features shown in Table \ref{tab:featuresxstage}, except \textit{Asym. Sim. - Title}. 

\noindent \textbf{\textit{CNN}:} we replace the \textit{Template} sentence similarity feature (i.e., \textit{Sent2Vec}) with \textit{CNN}. This means the embeddings representing the title of Q\&A pair generated by \textit{Sent2Vec} are replaced with embeddings generated using \textit{CNN}.

\noindent \textbf{CRAR - CNN and CRAR (Sent2Vec):} besides CRAR, which uses \textit{Sent2Vec} as the sentence similariry component, we construct another version (CRAR - CNN) where the sentence similarity component uses the CNN instead.

\begin{table}[t]
\centering
\caption{Performance of CRAR (combined score) and other  baselines derived from individual scores for ranking threads in terms of Hit@K, MRR@K, MAP@K, and MR@K, for K=10}
\label{tab:baselinesThreads}
\begin{tabular}{lcccc}
\hline
\textbf{Baseline Approach} & \textbf{Hit} & \textbf{MRR} & \textbf{MAP} & \multicolumn{1}{c|}{\textbf{MR}} \\ \hline
TF                                             & 0.67 & 0.39 & 0.35 & 0.18 \\  
Sent2Vec                                       & 0.68 & 0.35 & 0.32 & 0.16 \\
CNN                                            & 0.72 & 0.37 & 0.33 & 0.15 \\ 
Asym.Sim.Body+Ans.Body              & 0.72 & 0.44 & 0.40  & 0.16 \\  
Asym.Sim.Title                             & 0.79 & 0.42 & 0.37 & 0.22 \\
CRAR Without TF                                & 0.82 & 0.56 & 0.50 & 0.22 \\
CRAR Without Sent2Vec/CNN                      & 0.84 & 0.59 & 0.51 & 0.26 \\
CRAR Without Asym.Sim.Body+Ans.Body & 0.84 & 0.60 & 0.50 & 0.24 \\
CRAR Without Asym.Sim.Title               & 0.86 & 0.58 & 0.50 & 0.24 \\
CRAR - CNN                               & 0.86 & 0.58 & 0.51 & 0.26 \\
CRAR (Sent2Vec)   & 0.89         & 0.62         & 0.53         & 0.26 \\ \hline
\end{tabular}
\end{table}

Since RQ \#2 has focus only on thread features, we keep the answer features weights for the 11 baselines. Then, like in RQ \#1, we run the 11 baselines to search for relevant answers (Section \ref{online}) against the testing set queries. The results as shown in Table~\ref{tab:baselinesThreads}. We notice that when isolated, the \textit{CNN} feature outperforms the \textit{Sent2Vec}. However, when combined with the other thread features, \textit{Sent2Vec} shows better results for the four metrics (i.e., Top-K Accuracy,  Mean Average Precision, Mean Recall, and Mean Reciprocal Rank). The same situation happens with the other features. For instance, TF when isolated has worse performance compared to \textit{Asymmetric Similarity}. But with all features combined (CRAR), the absence of \textit{TF} produces the worst baseline, i.e, the \textit{hit} decays from 0.89 in \textit{CRAR baseline} to 0.82 on \textit{CRAR without TF} baseline.

\vspace{.5em}
\begin{center}
\fbox{\begin{minipage}{32em}
\textbf{RQ2 Answer:} In isolation, CNN baseline outperforms Sent2Vec baseline (lines 2 and 3 of Table~\ref{tab:baselinesThreads}). So, we would expect that when combined the CNN could deliver a better integrated approach, but surprisingly, Sent2Vec worked better together with the other combined features, i.e., baselines \textit{CRAR (Sent2Vec)} and \textit{CRAR - CNN} in Table~\ref{tab:baselinesThreads}. A possible reason is because Sent2Vec captures posts that the other features could not capture, suggesting to be more complementary to the other features than CNN. 
All individual features play a role for improving CRAR retrieval performance because if any of them are removed CRAR performance is decreased.

\end{minipage}}
\end{center}
\vspace{.5em}

\textbf{RQ3:} \textit{How the individual answers features can influence the ranking of candidate answers compared to the features combined (CRAR)?}

We investigate how each of the four individual answer features (i.e., Table~\ref{tab:featuresxstage}) influence the ranking of candidate answers. Similar to RQ \#2, we analyze their individual performance against their combination (CRAR) and the respective \textit{CRAR Without $<$Feature$>$}. For this, we construct four baselines where in each one we keep the weights associated to the thread features and set the weight of the answer features to zero, except the weight for the representing feature. For example, for the baseline \textit{Top Method}, we set the weight of all answer features to zero, except the weight associated to the \textit{Top Method}. The \textit{CRAR Without $<$Feature$>$} does the inverse, setting the weight to zero only for \textit{Top Method}, maintaining the proposed weights (Table~\ref{tab:featuresxstage}) in the other answer features. We then run the eight baselines to search for relevant answers (Section \ref{online}) against the testing set queries. 

The results are shown in Table~\ref{tab:baselinesAnswers}. The removal of any of those features impacts CRAR metrics, indicating that their combination are  effective for the retrieval performance. 

\begin{table}[t]
\centering
\caption{Performance of CRAR (combined score) and other  baselines derived from individual scores for ranking \textbf{answers} in terms of Hit@K, MRR@K, MAP@K, and MR@K, for K=10}
\label{tab:baselinesAnswers}
\begin{tabular}{lcccc}
\hline
\textbf{Baseline Approach} & \textbf{Hit} & \textbf{MRR} & \textbf{MAP} & \multicolumn{1}{c|}{\textbf{MR}} \\ \hline
Top Method                         & 0.61         & 0.35         & 0.31         & 0.11 \\
TF-IDF                             & 0.63         & 0.31         & 0.30         & 0.15 \\
Thread Score                       & 0.63         & 0.33         & 0.33         & 0.17 \\
Asymmetric Similarity              & 0.68         & 0.34         & 0.31         & 0.13 \\
CRAR Without TF-IDF                & 0.79         & 0.46         & 0.42         & 0.19 \\
CRAR Without Asymmetric Similarity & 0.82         & 0.53         & 0.48         & 0.24 \\
CRAR Without Thread Score         & 0.84         & 0.48         & 0.44         & 0.21 \\
CRAR Without Top Method           & 0.84         & 0.54         & 0.47         & 0.24 \\
CRAR (Sent2Vec)                    & 0.89         & 0.62         & 0.53         & 0.26 \\ 
\hline
\end{tabular}
\end{table}

\vspace{.5em}
\begin{center}
\fbox{\begin{minipage}{32em}
\textbf{RQ3 Answer:} The baseline containing the four features of the answers combined (i.e., CRAR with Sent2Vec) outperforms the four baselines based on each isolated feature by a significant margin for the four adopted metrics. Moreover, the removal of any of them would decrease the retrieval performance.
\end{minipage}}
\end{center}
\vspace{.5em}

\textbf{RQ4:} \textit{How does CRAR perform compared the state-of-art CROKAGE in retrieving relevant answers for given programming tasks?}
To answer this question, we rely on two approaches. The first one replicates CROKAGE in similar conditions as CRAR using the available ground-truth and compare the respective retrieval performance metrics.
The second approach relies on a user study, where we sample rated questions (in a 5-star score) from the CROKAGE public service\footnote{http://isel.ufu.br:9000}. Then, we ask independent users to blindly rate a sample with best and worst rated CROKAGE's answers.

\textbf{Ground-truth comparison.}
We replicate CROKAGE~\cite{silva2019recommending} using the artifacts provided in the respective companion site. CROKAGE synthesizes answers after retrieving the most relevant answers for a given query. Since we are only interested in its retrieval mechanism, we disregard its answers' synthesis in our experiments. Furthermore, CROKAGE relies on external tools to generate API classes. We do not reproduce the external tools. Instead, we leverage a cache provided by the authors of CROKAGE containing the classes produced by these tools for all the queries of the testing set. We then run CROKAGE to search for relevant answers (Section \ref{online}) against the testing set queries. The results are shown in Table~\ref{tab:CROKAGEvsCRAR}.

\begin{table}[h]
\centering
\caption{Performance of CRAR compared to the state-of-art CROKAGE in terms of Hit@K, MRR@K, MAP@K, and MR@K, for K=10}
\label{tab:CROKAGEvsCRAR}
\begin{tabular}{lcccc}
\hline
\textbf{Baseline Approach} & \textbf{Hit} & \textbf{MRR} & \textbf{MAP} & \multicolumn{1}{c|}{\textbf{MR}} \\ \hline
CROKAGE                    & 0.82         & 0.51         & 0.47         & 0.20                             \\
CRAR      & 0.89         & 0.62         & 0.53         & 0.26                             \\ \hline
\end{tabular}
\end{table}

In order to verify whether this superiority is significant, we run the non-parametric Wilcoxon signed-rank test~\cite{wilcoxon1945individual} on paired data of the two approaches. The test attests significant difference (i.e., p-values $<$ 0.05) between CROKAGE and CRAR for Mean Reciprocal Rank and Mean Recall with effect sizes of 0.26 (small) and 0.34 (medium) respectively, calculated with $r=Z/\sqrt{n}$ \cite{fritz2012effect}. That is, compared to CROKAGE, CRAR misses less relevant answers than CROKAGE (i.e., better recall). Furthermore, CRAR recommends relevant answers closer to the top positions (i.e., better reciprocal rank).




All the experiments were conducted over a server equipped with Intel\textregistered{} Xeon\textregistered{} at 3.1 GHz on 86 GB RAM, 12 cores, and 64-bit Linux Mint Cinnamon operating system. 
CRAR spends 44 seconds (mean after 10 executions, with standard deviation = 4.22) to retrieve the results for the 57 queries of the test set (i.e., less than one second per query).

\textbf{Qualitative user study.} To evaluate how CRAR and CROKAGE compare in providing comprehensive solutions to developers, we designed a user study with other queries asked by developers worldwide. CROKAGE is a publicly available service\footnote{http://isel.ufu.br:9000}. Users can send a query, obtain the answers and anonymously rate the answers using a 5-star scale. For this study, we randomly select 40 queries made in CROKAGE site, of which half of them (20 queries) were classified by  users  with 1 or 2 stars. The 20 other queries were rated with 5 stars. Then for each query, we collected the top-1 answer from CRAR and CROKAGE, calling them as Tool A and Tool B, respectively, for a blind assessment. Then, \red{we sought for volunteers in the Post-Graduate Program in Computer Science at Federal University of Uberlândia, and  three independent members accepted to participate: one postdoctoral researcher, one PhD candidate and, one Master student} with more than 10 years experience in Java. They were asked to provide a Likert value from 1 to 5 for the relevance of the suggested Stack Overflow answer for the each query, in each solution (CRAR and CROKAGE). We used Krippendorff's $\alpha$ reliability  coefficient \cite{Krippendorff2011ComputingKA} to verify the agreement between the three independent users. After one round, where users could review their grades in cases where the other two had  a difference of two or more points difference, we obtained $\alpha = 0.693$ for CRAR evaluation and $\alpha = 0.694$ for CROKAGE evaluation, which are acceptable agreements \cite{krippendorff04}, for evaluating the assessment.

\begin{figure}
     \centering
     \begin{subfigure}[b]{0.5\textwidth}
         \centering
         \includegraphics[width=\textwidth]{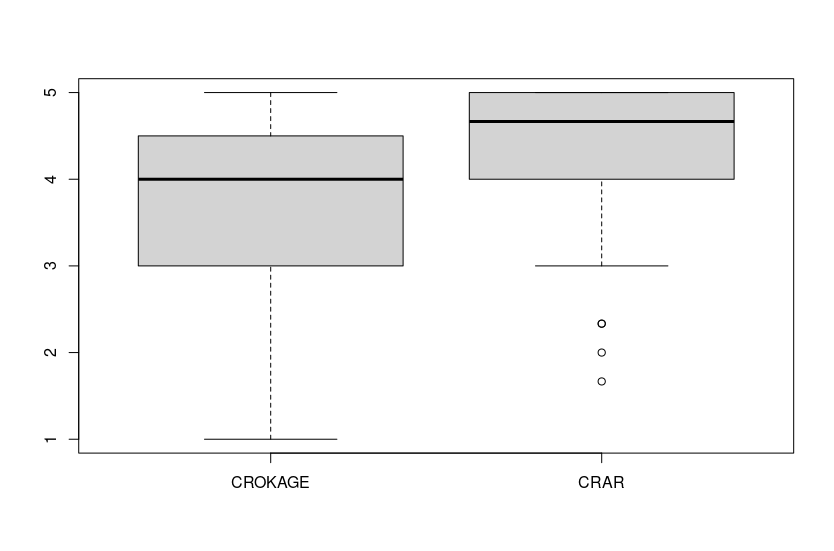}
         \caption{Relevance for all 40 Answers}
         \label{boxplot1}
     \end{subfigure}
     \newline
     \begin{subfigure}[b]{0.46\textwidth}
         \centering
         \includegraphics[width=\textwidth]{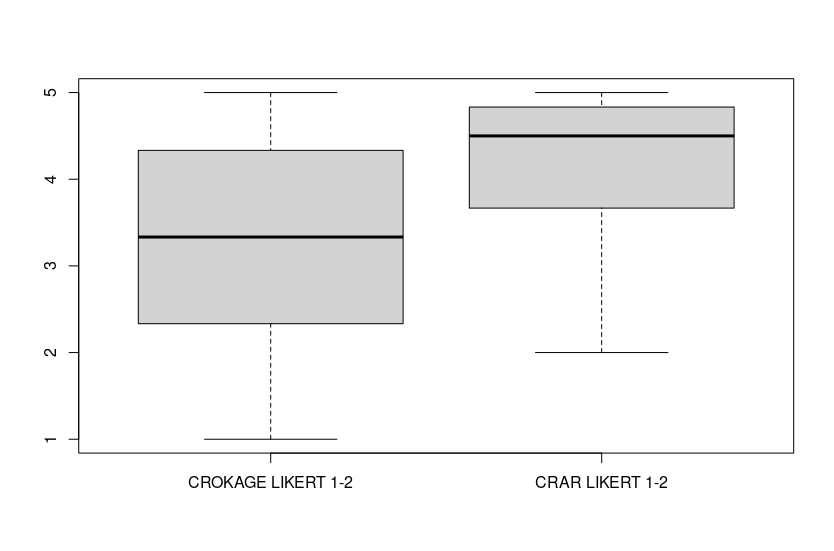}
         \caption{Relevance for Answers with likert 1-2.}
         \label{boxplot2}
     \end{subfigure}
     \hfill
     \begin{subfigure}[b]{0.46\textwidth}
         \centering
         \includegraphics[width=\textwidth]{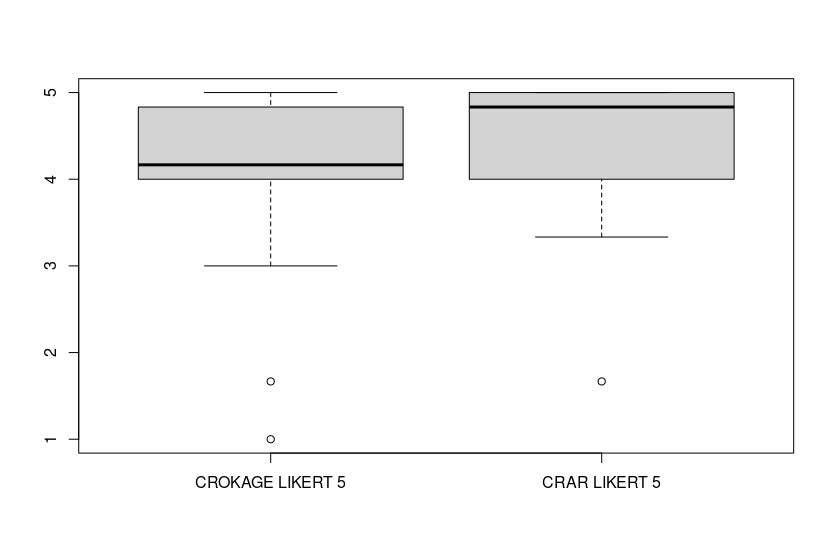}
         \caption{Relevance for Answers with likert 5.}
         \label{boxplot3}
     \end{subfigure}
        \caption{Box plots of Relevance of the Suggested CRAR and CROKAGE solutions.}
        \label{fig:boxplot}
\end{figure}

The Figure~\ref{boxplot1} represents the participants evaluation in all 40 queries. We run Wilcoxon signed-rank test, and CRAR is statistically better than CROKAGE with a confidence level of 95\% (p-value $<$ 0.05), with medium effect size 0.484. For a more detailed analysis of the data, we show  in Figure~\ref{boxplot2}, a boxplot for only the 20 queries with 1-2-stars from CROKAGE service assessment. We observed a better rate for CRAR answers, with statistically significant difference using the Wilcoxon signed-rank test  (p-value $<$ 0.05). In Figure~\ref{boxplot3}, we show the boxplot for the other 20 queries rated with 5-star. Although CRAR seems to have performed better, we could not find statistical significance using the Wilcoxon signed-rank test (p-value $>$ 0.05). This is an expected result because CRAR had already a very good rating from the external worldwide developers.

\noindent
\begin{center}
\fbox{\begin{minipage}{31em}
\textbf{RQ4 Answer:} Compared to the state-of-art CROKAGE,  CRAR shows improvements for the four adopted metrics in the retrieval of relevant answers for given programming tasks. CRAR outperforms CROKAGE in terms of Top-K Accuracy, Mean Average Precision, Mean Recall, and Mean Reciprocal Rank for K=10 by 7\%, 6\%, 6\%, and nominally 0.11, respectively. Moreover, a qualitative user study has shown that CRAR is at least as good as CROKAGE on queries that CROKAGE has already  good retrieval performance, and has a significant better performance on queries where CROKAGE presents limitations.
\end{minipage}}
\end{center}
\vspace{.5em}

\section{Discussion}\label{sec:discussion}
\begin{figure}
\centerline{\includegraphics[width=0.75\textwidth]{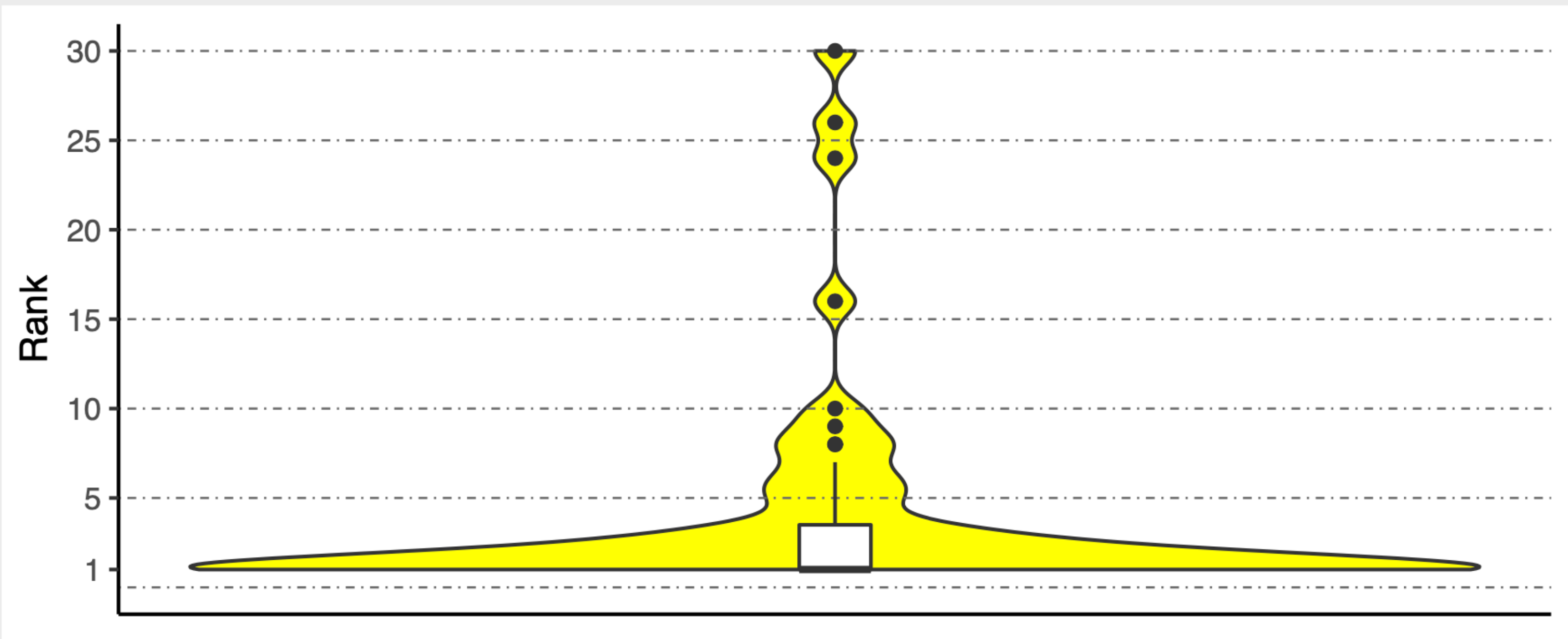}}
\caption{Best rank distribution for the test set queries.}
\label{fig:bestRank}
\end{figure}

The quantitative results have shown the effectiveness of CRAR in providing adequate answers for the queries available in the proposed test set. Figure~\ref{fig:bestRank} shows the distribution of the best ranked correct answer provided by CRAR for the 57 queries available in the test set. Two queries from the test set had no correct answer in the top-100, and could not be plotted.  We can observe that half of the answers are ranked at top-1. 

We manually analyzed those queries to understand important characteristics of the queries and the successful result of CRAR in these cases. We observed that in general there is high lexical similarity between the query and the returned Q\&A pair. So, since our starting point is based on lexical similarity provided by BM25, the chances of good results are high. Interestingly, we  observe that, in several cases, the lexical similarity between the query and the returned question is not high. However, the answer similarity tends to be high, showing that the global strategy to select relevant threads and then fine tuning the selection of answers seems to have contributed to the successful result. 
 
Moreover, we also manually analyzed the queries to understand characteristics that has led to the worse results of CRAR. There are different characteristics that help to explain why CRAR did not achieve better results on those cases.

One of those characteristics is that although there is high lexical similarity between the query and the Q\&A pair, the intrinsic semantics of the solution does not match with the intent of the query. For instance, for the query \textit{``List XML element Attributes"}, the correct answer is in the $5^{th}$ position, whose question title is \textit{``Get the list or name of all attributes in a XML element"}. However, in the $1^{st}$ position, CRAR returns an answer whose question title is \textit{``How to get both the attributes and the data list with XmlList"}. In another example, for the query \textit{``Java Program to display first 100 prime numbers"} CRAR returns as top-1 a solution with high textual similarity but with a subtle semantic difference: a program to display primes up to 100.

Another characteristic is that more conceptual queries, such as \textit{``How do I extend classes in Java?"}, may lead to answers with high textual similarity, but with much more specific content, thus not meeting adequately the query intent of a more conceptual answer. For instance, the answer for  \textit{``How to unit test abstract classes extend with stubs?''} is returned, which provides a high lexical similarity, but does not meet the required question semantics. In this case, we could see that our proposed semantic features, Sent2Vec and CNN, still were not able to capture the actual semantics of the question, and thus this kind of feature deserves more investigation.

Although the antonyms feature has contributed to CRAR, there are more subtle antonyms that are intrinsic to the semantics of a function. For instance, for the query
\textit{``Convert URL query String to Map"}, while the correct solution points to a function that parses a URL and generate respective map, the best ranked solution encodes a URL from a map of parameters.
Moreover, a subtle difference in the meaning of terms can lead to inadequate answer, such as in the query \textit{``Java Program to Calculate average of numbers using Array"}. In this case, the first solution inadequately provides an algorithm using an ArrayList instead of array, because the solution includes the compound noun \textit{``array list"}.

Summing up, although the semantic similarity scores adopted by CRAR have contributed to improve the results, most of the inadequate results could still be credited to the inability of CRAR to extract more semantics from the Q\&A pairs to capture the intent of the query more effectively. 

\section{Threats to validity}\label{sec:threats}

\textbf{Threats to internal validity} are related to the biases and experimental errors. In this work we replicate a state-of-art tool~\cite{silva2019recommending} and customize several other baselines to produce the results. In order to mitigate the bias, we used the same ground truth and the same testing set provided the state-of-art tool to run all baselines.  Nonetheless, the queries from the ground-truth has been independently extracted from third-party popular sources, it may not be representative of the universe of queries issued by developers. Furthermore, we double check our baseline implementations to assure that they do not contain implementation errors. 

\textbf{Threats to external validity} relates to the dictionary of antonyms and the generalizability of our results. First, the dictionary of antonyms could contain bias and errors. We mitigate the bias by merging three different external repositories. But since we do not check all antonyms (i.e., around 14k), errors in the resulting list can exist. Second, regarding the generalizability of our results, our baselines are limited to the Java language the same way as the state-of-art tool~\cite{silva2019recommending}. However, we do not consider this limitation as a threat since Java is a popular language among developers. The Stack Overflow dataset used to test all baselines contains around 4.1M Java posts (questions and answers). Furthermore, our baselines could be easily adapted to support other languages, as long as they can access a Stack Overflow dataset.  

\textbf{Threats to construct validity} relates to the suitability of our evaluation metrics. We adopt the same performance metrics used by the state-of-art tool~\cite{silva2019recommending}: Top-K Accuracy, Mean Reciprocal Rank, Mean Average Precision and Mean Recall. These metrics have also been adopted by several related works in the software engineering literature~\cite{rahman2019supporting,rahman2016rack, rahman2018effective,rahman2017strict,huang2018api,xu2017answerbot}, confirming little or no threat to our construct validity.


\section{Related Work}\label{sec:relatedWord}

A number of studies on Information Retrieval leverages the crowd knowledge to help developers in software development~\cite{cao2018toward,diamantopoulos2015employing,ponzanelli2014mining,rahman2018effective,cambronero2019deep,xu2017answerbot,ye2016word,huang2018api,gu2018deep}. Some works~\cite{diamantopoulos2015employing,ponzanelli2014mining} employ traditional IR techniques such as TF-IDF 
to recommend relevant discussions from Stack Overflow for a given context. Differently from our work whose input is a natural language query representing a programming task, their context carry not only textual information but also code. 

Information Retrieval has been also employed over Stack Overflow to assist documentation~\cite{cao2018toward,treude2016augmenting,rahman2015recommending}. Cao et al.\cite{cao2018toward} combine TF-IDF with machine learning to generate traceability links between code and documentation. Similarly, Treude et al.\cite{treude2016augmenting} employs machine learning to link API documentation with sentences from Stack Overflow. Rahman and Roy~\cite{rahman2015recommending} combine PageRank~\cite{brin2017anatomy} with heuristics to mine insightful comments from Stack Overflow to complement code. 

Since the keyword based approaches falls short to capture the semantics of words, studies have been leveraging embeddings to mine knowledge from the crowd~\cite{ye2016word,xu2017answerbot,huang2018api,rahman2018effective,silva2019recommending}. Rahman and Roy~\cite{rahman2018effective} combine embeddings with pseudo-relevance feedback~\cite{rahman2017improved,nie2016query,haiduc2013automatic} to identify relevant API classes from Stack Overflow to assist code search through reformulation of queries. Ye et al.\cite{ye2016word} improve a traditional lexical similarity measure~\cite{mihalcea2006corpus} and devise a more advanced mechanism called asymmetric relevance to calculate the similarity of two bag-of-words considering their semantics. 

Several  works~\cite{xu2017answerbot,huang2018api,silva2019recommending} employ the formula devised by Ye et al.\cite{ye2016word} to mine the crowd knowledge to provide solutions for developer issues. AnswerBot~\cite{xu2017answerbot} and BIKER~\cite{huang2018api} use the asymmetric relevance to retrieve the most relevant questions for a given query and then, out of their answers, generate solutions for the query. CROKAGE~\cite{silva2019recommending} instead, applies the asymmetric relevance in combination with BM25~\cite{robertson1994some} function to retrieve relevant answers for a query. Our work is similar to CROKAGE since their solutions, unlike  BIKER and AnswerBot, contains both code and explanations for a wide range of APIs contained in the answers of Stack Overflow.

\section{Conclusion and Future Work} \label{sec:conclusion}

In this work, we propose a novel approach called CRAR that recommends solutions for developers' programming tasks. The solutions are in form of answers obtained from Stack Overflow containing code examples and explanations. Unlike a previous state-of-art tool though, CRAR first retrieves the most important threads for a given programming task, and then, extracts relevant answers from them. For this, CRAR employs a combination of a well known Information Retrieval function and a series of features.

We found that among the investigated features asymmetric similarity  plays an important role. Considering the semantic features, we found that
although  CNN  embedding  performs  better  in  isolation compared  to  Sent2Vec,  the  latter  produces  a  better  combined  score  for CRAR, suggesting that Sent2Vec is more complementary to the other features.   
 Several combinations for the antonym feature were evaluated and the one that resulted in better metrics was the one that  discards  answers (\textit{ANS}) containing antonyms for the nouns (\textit{NN}) in the query. Although, the use of the antonym feature had shown improvement on the overall metrics, such improvement was not statistically significant.
 The use of the social feature \textit{Thread Score} was shown to strongly impact on the ranking of candidate answers, better than \textit{Question Score} and \textit{\#Answers}. 
 Finally, we show that the combination of the proposed features performed better than those features in isolation.

Finally, we show that CRAR outperforms the state-of-art tool to recommend answers for programming tasks by a statistical significant margin. 

As future work, we plan to implement CRAR in the form of an online tool to assist developers with their daily programming tasks and investigate new solutions for neural information retrieval based on deep learning architectures.

\section*{Acknowledgments}
This research was supported in-part by a Canada First Research Excellence Fund (CFREF) grant coordinated by the Global Institute for Food Security (GIFS). We  thank the Brazilian funding agencies, CAPES, CNPq and FAPEMIG for partially supporting this research.

\bibliographystyle{elsarticle-num}
\bibliography{biblio} 

\end{document}